\documentclass[conference]{IEEEtran}
  \pdfoutput=1\relax                   
  \usepackage{graphicx}                
  \DeclareGraphicsExtensions{.pdf,.png,.jpg,.jpeg} 
  \usepackage{graphicx}                
  \DeclareGraphicsExtensions{.eps}     

\IEEEoverridecommandlockouts
\graphicspath{{figures/}{pictures/}{images/}{./}} 
\usepackage{amsfonts}
\usepackage{microtype}                 
\PassOptionsToPackage{warn}{textcomp}  
\usepackage{textcomp}                  
\usepackage{amsmath}
\usepackage{mathptmx}                  
\usepackage{times}                     
\usepackage{cite}                      
\usepackage{svg}
\usepackage{float}
\usepackage{subcaption}
\usepackage{caption}
\usepackage{tabu}                      
\usepackage{booktabs}                  
\usepackage{url}
\usepackage{eso-pic}
\title{Lightweight CFR-Based Modulation Adaptation in a Real-Time MIMO-OFDM SDR Testbed}

\author{
\IEEEauthorblockN{Luca Borst, Maryam Ansarifard, Ankith Vinayachandran, Kishor C. Joshi, George Exarchakos}

\IEEEauthorblockA{Department of Electrical Engineering, Eindhoven University of Technology, Eindhoven, The Netherlands\\
Emails: L.Borst@student.tue.nl, \{M.Ansarifard, A.Vinayachandran, K.C.Joshi, G.Exarchakos\}@tue.nl}
}
\begin{document}
\AddToShipoutPictureFG*{%
  \AtPageUpperLeft{%
    \put(0,-10){%
      \makebox[\paperwidth][c]{%
        \scriptsize\itshape
        This work has been submitted to the IEEE for possible publication.
        Copyright may be transferred without notice, after which this version may no longer be accessible.
      }%
    }%
  }%
}
\maketitle

\begin{abstract}
Conventional link adaptations typically rely on scalar link-quality indicators such as signal-to-noise ratio (SNR), while richer channel state information (CSI) can improve adaptation at higher processing complexity. This paper investigates a compact alternative for modulation selection in a real-time multiple-input multiple-output orthogonal frequency-division multiplexing (MIMO-OFDM) system, using channel frequency response (CFR) magnitude descriptors. A dataset of 87{,}817 over-the-air (OTA) samples is collected using a USRP-based testbed, with CFR measurements extracted at the base station (BS) from received uplink pilots. Decision tree (DT), random forest (RF), and $k$-nearest neighbours (KNN) classifiers are evaluated using BS-side SNR, CFR features, and their combination. SNR-only classifiers achieve 35\%--42\% test accuracy, whereas CFR-only features achieve 73.6\%, 81.4\%, and 80.0\% for DT, RF, and KNN, respectively. CFR-based performance is maintained near the 10\% BLER reliability thresholds, with RF reaching 82.8\%. A depth-7 DT with 123 leaves is further integrated into the LabVIEW C~Node for real-time inference. The results show that compact BS-side CFR descriptors provide more discriminative information than the available scalar BS-side SNR while remaining suitable for lightweight SDR implementation.
\end{abstract}
\begin{IEEEkeywords}
link adaptation, adaptive modulation, machine learning, MIMO-OFDM, software-defined radio
\end{IEEEkeywords}

\section{INTRODUCTION}

Link adaptation is a key mechanism for maintaining reliable and spectrally efficient communication over time-varying wireless channels. By adapting transmission parameters to the instantaneous channel conditions, a wireless system can balance robustness against achievable data rate. This becomes particularly important in multiple-input multiple-output orthogonal frequency-division multiplexing (MIMO-OFDM) systems, where the propagation channel varies not only over time but also across transmit--receive antenna pairs and OFDM subcarriers. Multipath propagation, frequency-selective fading, interference, and path loss can therefore affect different parts of the transmitted signal differently. Modulation adaptation addresses this trade-off by selecting a more robust low-order modulation under unfavourable channel conditions and a higher-order modulation when the channel can support a larger data rate.

Link-adaptation decisions require information about the propagation channel. In frequency-division duplexing (FDD) systems, channel state information (CSI) is generally estimated at the receiver from reference or pilot symbols and conveyed to the transmitter through channel-quality feedback. In time-division duplexing (TDD) systems, however, uplink and downlink propagation occur in the same frequency band at different time instances. Under the channel-reciprocity assumption, uplink pilot measurements can therefore be used at the base station (BS) to obtain a representation of the corresponding downlink channel, reducing the need for explicit receiver-side CSI feedback.

Conventional link adaptations commonly relies on scalar link-quality indicators, such as signal-to-noise ratio (SNR), signal-to-interference-plus-noise ratio (SINR), or channel quality indicator (CQI), to a suitable transmission configuration using predefined thresholds. Outer-loop link adaptation (OLLA) further adjusts these thresholds according to ACK/NACK feedback in order to compensate for prediction errors and changing channel conditions \cite{Blanquez-Casado2016,Mota2019,mohan2019}. Such approaches provide low-complexity decision mechanisms and are well suited to practical implementations. However, reducing the channel condition to a scalar metric does not explicitly retain the frequency-selective structure present in a MIMO-OFDM channel. Consequently, two different channel realisations with similar average SNR may exhibit different channel-frequency responses across the OFDM bandwidth, potentially resulting in different modulation reliability.

\subsection{Related Work}

Machine learning (ML) has increasingly been investigated as an alternative to fixed threshold-based link adaptations \cite{bobrov2021massive}. Several approaches retain conventional scalar link-quality indicators as model inputs. For example, contextual or supervised learning methods have used SNR \cite{pivoto2026contextual}, SINR, CQI, received-signal-strength information, and ACK/NACK \cite{wiesmayr2025salad} observations to predict suitable modulation and coding schemes. These methods improve the mapping between conventional link-quality measurements and transmission decisions, but the channel is still represented either through a small number of scalar indicators or indirectly through receiver feedback.

A different line of work exploits richer CSI representations. In \cite{an2023ml}, for example, a feedback-free ML-based link-adaptation framework uses the full complex channel matrix as input to a CNN-LSTM architecture for MCS prediction, achieving 92.5\% test accuracy and outperforming threshold-based baselines. Such approaches demonstrate that high-dimensional CSI contains information that can be useful for link adaptation beyond conventional scalar metrics. However, processing the full complex CSI matrix increases the dimensionality of the learning problem and typically requires more complex model architectures and inference operations.

ML-based \emph{modulation-only} adaptation has also been considered independently of coding-rate optimisation. Decision-tree-based modulation selection was investigated in \cite{ismail2025cognitive}, while a random-forest-based modulation-order selection approach was proposed in \cite{shao2024adaptive}. These works demonstrate that modulation adaptation can be treated as an independent ML classification problem. However, they were primarily evaluated in simulated narrowband or hybrid FSO/RF environments rather than in a real-time MIMO-OFDM SDR implementation.

The existing literature therefore presents two main design directions: low-complexity link adaptation based on scalar channel-quality indicators, and richer CSI-based approaches that retain substantially more channel information at the cost of increased dimensionality and implementation complexity. Comparatively less attention has been given to the intermediate design point between these two extremes; exploiting the frequency-dependent channel information available in the channel frequency response (CFR), while compressing it into a small number of descriptors suitable for lightweight real-time inference. The CFR describes the complex-valued gain per OFDM subcarrier, representing the channel in the frequency domain, whereas CSI also covers time-domain or spacial characteristics \cite{Tse2005}. 

This work investigates this intermediate design point using an over-the-air (OTA) MIMO-OFDM testbed implemented with NI USRP hardware and the LabVIEW Communications framework \cite{NI_MIMOAppFW2018}. Under TDD channel reciprocity, uplink pilot measurements are processed at the BS to obtain the CFR. Instead of providing the complete complex CFR across all subcarriers to the learning model, its magnitude response is compressed into four statistical descriptors (mean, standard deviation, range, and slope), representing the average channel magnitude and its variation across frequency. These descriptors are then used for ML-based modulation selection.

The study considers modulation adaptation independently of channel-coding-rate optimisation in order to isolate the effect of the channel representation on the modulation-selection problem. Three supervised classifiers, a decision tree (DT), random forest (RF), and $k$-nearest neighbours (KNN), are evaluated using three input configurations: BS-side SNR only, CFR magnitude descriptors only, and their combination. The resulting performance is evaluated using OTA measurements, including samples close to the modulation reliability boundaries. Finally, a complexity-constrained DT is translated into C-based if--else rules and integrated into the real-time LabVIEW processing chain.


The main contributions of this work are summarised as follows:
\begin{itemize}
    \item An OTA experimental evaluation of CFR-based modulation selection in a real-time MIMO-OFDM SDR testbed using compact CFR magnitude descriptors rather than the full complex CSI matrix.

    \item A systematic comparison of BS-side SNR, CFR-derived features, and their combination across DT, RF, and KNN classifiers, including evaluation close to the modulation reliability boundaries.

    \item BS-side CFR-based classifiers outperform SNR-only classifiers (73.6\%--81.4\% vs. 35\%--42\% test accuracy), with performance maintained or improved near the 10\% BLER reliability thresholds, reaching 82.8\% for the RF CFR+SNR classifier.

    \item A complexity-aware real-time implementation of a depth-7 DT with 123 leaves in the LabVIEW C~Node, enabling modulation inference using a bounded sequence of threshold comparisons without online retraining or GPU-based inference.
\end{itemize}


\section{SYSTEM MODEL}\label{system model}
\label{sec:system_model}
This section gives a system overview and describes the signal model and the experimental OTA setup used in this study. 

\subsection{System Overview}
\begin{figure*}[!t]
    \centering
    \includegraphics[width=\textwidth,
    trim={0mm 25mm 10mm 4mm},
    clip]{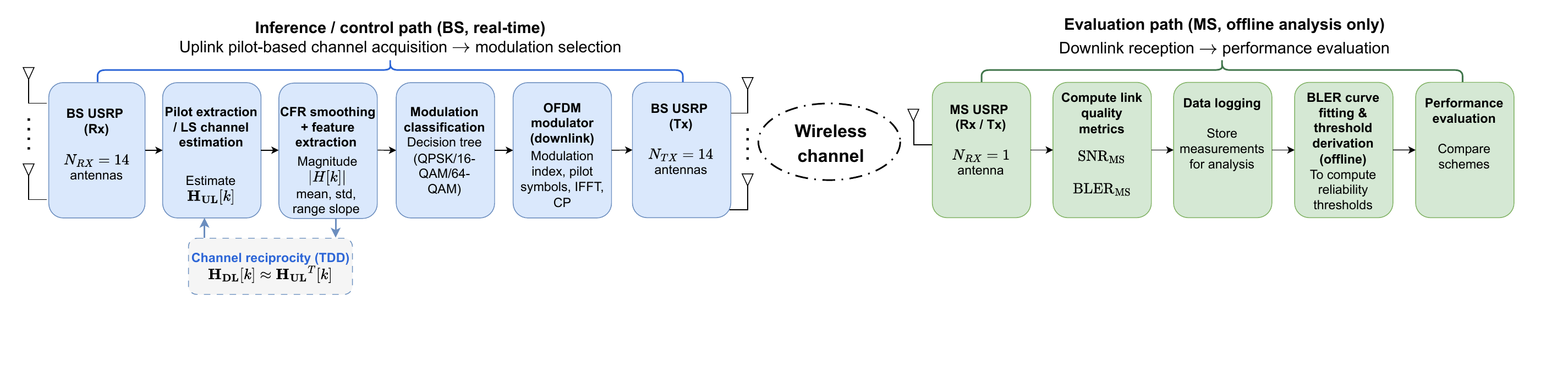}
    \caption{System architecture of the TDD MIMO-OFDM testbed. Uplink pilots transmitted by the MS are used for BS-side CFR estimation using a least-squares (LS) algorithm. Under channel reciprocity, CFR-derived features are provided to the deployed decision-tree classifier for downlink modulation selection. MS-side SNR and BLER are used only for measurement analysis and performance evaluation.}
    \label{fig:system-overview}
\end{figure*}

The testbed consists of a real-time MIMO-OFDM system implemented using NI USRP-2954 SDR hardware and the LabVIEW Communications framework. The system operates in a downlink configuration where $N_t$ and $N_r$ denote the number of antennas at the BS and the mobile station (MS), respectively. In this setup, $N_t=14$ and $N_r = 1$, resulting in a frequency-dependent channel matrix $\mathbf{H}[k] \in \mathbb{C}^{N_r \times N_t}$ (see Section~\ref{sec:MIMO_model}). This is physically implemented using 7 NI USRP-2954 SDRs forming the BS and 1 NI USRP-2954 SDR forming the MS, all connected to NI VERT2450 antennas. On the downlink, the BS performs modulation selection and OFDM waveform generation, including pilot insertion and cyclic-prefix addition. For channel acquisition, the MS transmits uplink pilot symbols that are received by the BS. The BS processes these pilots to estimate the uplink CFR and, under the TDD channel-reciprocity assumption, uses this estimate as a representation of the corresponding downlink channel. CFR magnitude descriptors are subsequently extracted and provided to the ML-based modulation selector. The selected modulation index is then passed to the downlink frame scheduler and OFDM modulator. The MS performs downlink OFDM demodulation and computes SNR and BLER measurements, which are logged for threshold derivation and performance evaluation but are not required as inputs to the CFR-based classifier during inference. An overview of this processing chain is shown in Fig.~\ref{fig:system-overview}.

\subsection{MIMO-OFDM Signal Model}
\label{sec:MIMO_model}
The wireless channel is characterised by the CFR across OFDM subcarriers \cite{goldsmith2005} as follows:
\begin{equation}
\mathbf{H} =
\begin{bmatrix}
    H[1] & H[2] & \cdots & H[N_{\mathrm{sub}}]
\end{bmatrix},
\end{equation}
where \(H[k] = |H[k]| e^{j\angle H[k]}\) denotes the complex-valued CFR of the \(k\)th subcarrier. Channel estimation for the proposed modulation-selection pipeline is performed at the BS using pilot symbols transmitted by the MS on the uplink. Let $\mathbf{H}_{\mathrm{UL}}[k]$ denote the estimated uplink CFR on the $k$th subcarrier. Under the TDD reciprocity assumption, the corresponding downlink propagation channel can be inferred from the uplink measurement, i.e.,
$\mathbf{H}_{\mathrm{DL}}[k]
\approx
\mathbf{H}_{\mathrm{UL}}^{T}[k],$
with the matrix transpose accounting for the reversal of transmitter and receiver roles. An automatic reciprocity calibration compensates for transceiver chain impact on the estimated channel, ensuring reciprocity \cite{NI_MIMOAppFW2018}.  The BS therefore uses the uplink-derived CFR as the channel representation for modulation selection. Channel estimation is performed via a LS algorithm at pilot subcarriers and applied to  MMSE-based precoder design at the BS \cite{Marzetta2016}. The number of subcarriers $N_{\mathrm{sub}}$ is equal to 1200, with a subcarrier spacing of 15~kHz. The system utilises 12 subcarriers per resource block (RB), resulting in $N_{\mathrm{RB}}=100$ resource blocks.

\subsection{OTA Measurement Setup}
\begin{figure}[h]
  \centering
  \includegraphics[width=0.6\columnwidth]{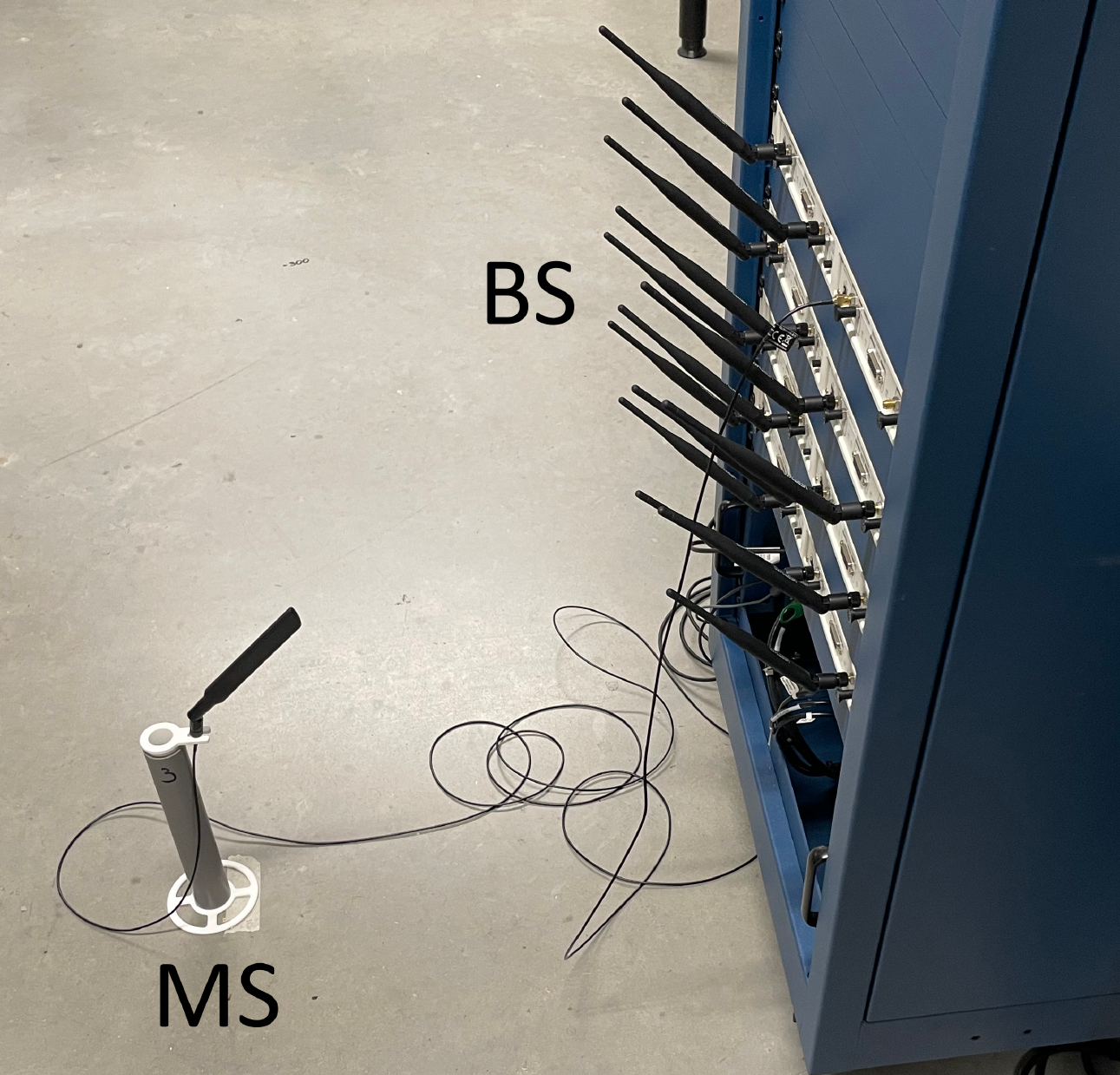}
  \caption{OTA measurement setup: NI USRP-2954 testbed with 
  14 BS antennas and 1 MS antenna at 60$~\mathrm{cm}$
  distance LoS.}
  \label{fig:OTA-setup}
\end{figure}

All experimental results are obtained using an OTA measurement configuration under controlled indoor conditions (Fig.~\ref{fig:OTA-setup}). The MS antenna is extended from the receiver host through a 5$~\mathrm{m}$, 50$~\Omega$ coaxial cable to simulate channel attenuation, positioned 60$~\mathrm{cm}$ from the testbed to create a line-of-sight (LoS) scenario. Communication is carried out at a carrier frequency of 4.2$~\mathrm{GHz}$ with a channel bandwidth of 20$~\mathrm{MHz}$. The MS remains stationary to ensure repeatable channel conditions during data acquisition.

\section{modulation adaptation methodology} \label{methodology}
This section describes data acquisition, processing, and the methods used to evaluate traditional and ML-based modulation adaptation under OTA conditions. 
\subsection{Data acquisition} \label{sec:data_acquisition}
Building on the OTA setup described in Section~\ref{sec:system_model}, measurements are collected under controlled channel conditions to construct the dataset used for classifier training and evaluation. For each modulation scheme, the BS transmit power is swept over 100 discrete levels, covering operating conditions from high-BLER regions to reliable transmission below the 10\% BLER threshold. At each power level, channel and link-quality measurements are recorded over a fixed observation interval of 30~s.

Two measurement streams are logged simultaneously. BS-side quantities, including the CFR and $\mathrm{SNR}_{\mathrm{BS}}$, are sampled at $f_{s,\mathrm{BS}}=10$~Hz, whereas the MS-side $\mathrm{SNR}_{\mathrm{MS}}$ and $\mathrm{BLER}_{\mathrm{MS}}$ are recorded at $f_{s,\mathrm{MS}}=1$~Hz. Because of the different sampling rates, each BS-side observation is associated with the temporally nearest MS-side measurement using timestamp synchronisation. This distinction is important for the proposed inference architecture. The BS-side quantities constitute the candidate inputs for modulation classification, while the downlink MS-side measurements are used only for (uncoded) BLER-curve fitting, reliability-threshold derivation, and offline performance analysis. The corresponding uplink BLER is not considered due to this study’s downlink-only scope. In particular, $\mathrm{SNR}_{\mathrm{MS}}$ and $\mathrm{BLER}_{\mathrm{MS}}$ are not provided to the deployed CFR-based classifier during real-time inference.

In addition to average channel quality, the CFR contains information about the variation of the channel response across OFDM subcarriers. Such frequency-selective behaviour arises from multipath propagation and is not explicitly represented by a single scalar SNR value \cite{Tse2005}. CFR-derived features are therefore extracted to provide a compact representation of both the average channel magnitude and its frequency-domain variation. Prior to feature extraction, the raw CFR magnitude is smoothed using a Savitzky--Golay filter \cite{Schafer2011,Milo2019}. The filter performs a local LS polynomial approximation that suppresses high-frequency measurement fluctuations while preserving the underlying shape of the frequency response \cite{Schafer2011}. A polynomial degree of $N=3$ and a window half-length of $M=49$ samples are used, corresponding to a total window length of $2M+1=99$ samples. 


For each modulation scheme, a logistic sigmoid is fitted to the
measured $\mathrm{BLER}_{\mathrm{MS}}$--$\mathrm{SNR}_{\mathrm{MS}}$
relationship using LS. The SNR corresponding to a target
BLER of 10\% is extracted from the fitted curve and used to define
the reliability and transition regions considered in Section~\ref{sec:data}. Before classifier training, constraint masks are applied to remove corrupted or physically implausible observations according to empirically selected bounds on the measured SNR and CFR magnitude statistics.

\subsection{Feature Engineering}\label{sec:feature_engineering}
Rather than using the full CFR across all OFDM subcarriers, the smoothed CFR magnitude is compressed into four descriptors: mean $\overline{|\mathbf{H}[k]|}$, standard deviation $\sigma_{|\mathbf{H}[k]|}$, range $\Delta_{|\mathbf{H}[k]|}$, and first-order LS slope $\hat{\alpha}_{|\mathbf{H}[k]|}$. These features capture the average channel magnitude, its variation, and the overall frequency-domain trend \cite{Tse2005}, while substantially reducing the input dimensionality.

Together with the BS-side SNR, these quantities form the set of candidate features available for modulation classification at the BS. The complete BS-side feature vector is defined as
\begin{equation}
    \mathbf{x}_{\mathrm{BS}} =
    \begin{bmatrix}
        \mathrm{SNR}_{\mathrm{BS}} &
        \overline{|\mathbf{H}[k]|} &
        \sigma_{|\mathbf{H}[k]|} &
        \Delta_{|\mathbf{H}[k]|} &
        \hat{\alpha}_{|\mathbf{H}[k]|}
    \end{bmatrix}^{\mathrm{T}}.
    \label{eq:feature_vector}
\end{equation}

To assess the contribution of the channel representation independently of the classifier architecture, three feature configurations are evaluated:
\begin{equation}
\begin{aligned}
    &\mathbf{x}_{\mathrm{SNR}}
        = \left[\mathrm{SNR}_{\mathrm{BS}}\right]^{\mathrm{T}}, \quad \mathbf{x}_{\mathrm{CFR}}
        = \left[
        \overline{|\mathbf{H}[k]|},
        \sigma_{|\mathbf{H}[k]|},
        \Delta_{|\mathbf{H}[k]|},
        \hat{\alpha}_{|\mathbf{H}[k]|}
        \right]^{\mathrm{T}},\\
    &\mathbf{x}_{\mathrm{CFR+SNR}}
        = \mathbf{x}_{\mathrm{BS}}.
\end{aligned}
\label{eq:feature_subsets}
\end{equation}

The MS-side quantities $\mathrm{SNR}_{\mathrm{MS}}$ and $\mathrm{BLER}_{\mathrm{MS}}$ are excluded from all classifier input vectors. As described in Section~\ref{sec:data_acquisition}, they are used only for BLER-curve fitting, reliability-threshold derivation, and offline performance analysis. Consequently, the evaluated feature sets contain only quantities available from the BS processing chain during inference.

The relative contribution of the individual features is analysed in Section~\ref{sec:feature_importance} to determine whether the classification performance is primarily associated with the average CFR magnitude or with its variation across frequency.

\subsection{Machine Learning-Based Modulation Selection}
Modulation selection is formulated as a three-class supervised
classification problem with QPSK, 16-QAM, and 64-QAM as the output
classes. The transmitted modulation is used as the class label, since BLER-based labelling leads to strongly imbalanced reliability regions. The fitted BLER thresholds are instead used for the transition-region analysis in Section~\ref{sec:boundary}.

DT, RF, and KNN classifiers are evaluated to compare classification performance and implementation complexity. The DT recursively partitions the feature space using feature-dependent thresholds and produces an interpretable sequence of if--else decisions \cite{good2023feature}. This structure is particularly suitable for embedded deployment because a bounded-depth tree can be translated directly into conditional C code. RF extends the tree-based approach by combining predictions from multiple trees trained on randomised subsets of the data and features, generally improving robustness at the expense of increased inference and implementation complexity \cite{James2023}. KNN provides a non-parametric comparison by assigning each observation to the most common class among its $k$ nearest training samples according to the selected distance metric \cite{halder2024enhancing}.

The dataset is split into 80\% training and 20\% test samples, with 5-fold cross-validation used for model selection and hyperparameter tuning. Each classifier is evaluated using the three BS-side feature configurations introduced in Section~\ref{sec:feature_engineering}: SNR-only, CFR-only, and CFR+SNR. This comparison isolates the effect of the channel representation from the choice of classifier. Feature importance is subsequently analysed for the tree-based models in Section~\ref{sec:feature_importance} to determine which BS-side measurements contribute most strongly to the predictions. RF and KNN serve as offline performance references, while DT is also considered for deployment because its bounded-depth decision rules can be translated directly into C if--else statements.

For real-time deployment, CFR extraction, smoothing, feature
computation, and DT inference are executed in the LabVIEW
processing chain at $f_{s,\mathrm{BS}}=10$~Hz. To avoid rapid modulation switching, the scheduler is updated only after the same modulation is predicted for three consecutive inference intervals, corresponding to 0.30~s.

\section{Experimental Results} \label{results}
This section presents the OTA measurement and offline classification results.

\subsection{Data Acquisition \& Channel Characterisation}
\label{sec:data}

\begin{figure}[tb]
    \centering
    \includegraphics[width=\columnwidth]{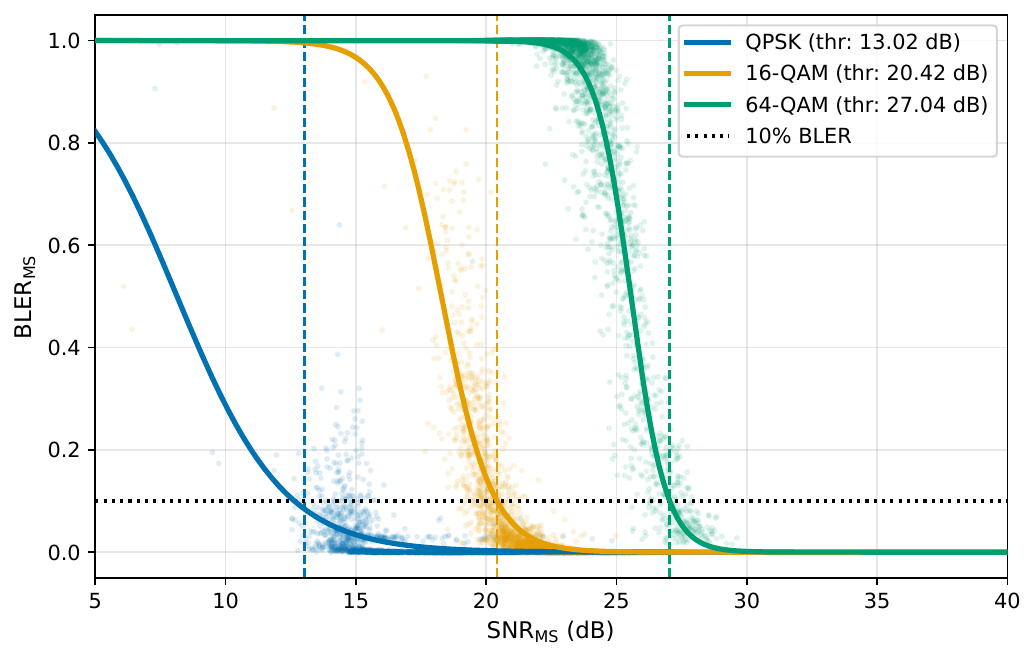}
    \caption{Measured MS-side BLER versus $\mathrm{SNR}_{\mathrm{MS}}$ for QPSK, 16-QAM, and 64-QAM under the 60~cm LoS configuration. The fitted sigmoid curves and corresponding 10\% BLER reliability thresholds are also shown.}
    \label{fig:bler_curves}
\end{figure}


After applying the constraint masks described in Section~\ref{sec:data_acquisition}, the final dataset contains 87{,}817 OTA observations, with 36.2\% QPSK, 28.2\% 16-QAM, and 35.5\% 64-QAM samples. 

Fig.~\ref{fig:bler_curves} shows the measured $\mathrm{BLER}_{\mathrm{MS}}$--$\mathrm{SNR}_{\mathrm{MS}}$ relationship and fitted sigmoid curves. The 10\% BLER reliability thresholds are 13.0~dB, 20.4~dB, and 27.0~dB for QPSK, 16-QAM, and 64-QAM, respectively, and are used in the transition-region analysis of Section~\ref{sec:boundary}.

Although the transmitted-modulation classes are relatively balanced,
the reliability regions are not: only 0.1\% of QPSK samples lie below its threshold, while only 5.3\% of 64-QAM samples lie above its threshold. This imbalance motivates the use of transmitted modulation rather than BLER-based labels for classifier training.

\begin{figure}[tb]
 \centering
 \begin{subfigure}{\columnwidth}
   \centering
   \includegraphics[width=0.8\columnwidth]{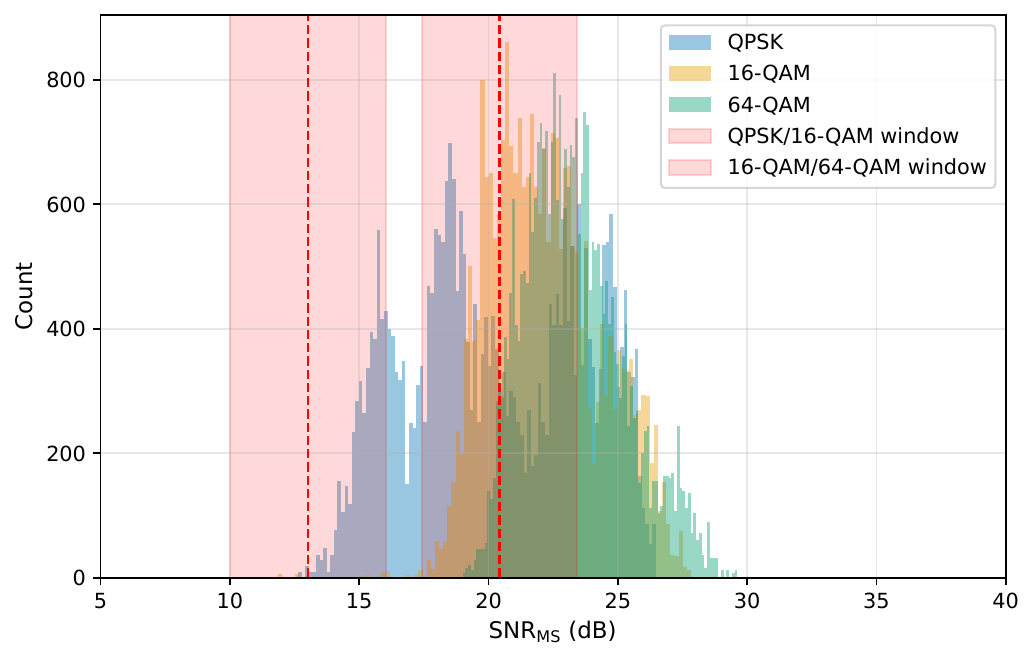}
   \caption{$\text{SNR}_{\text{MS}}$ distribution with $\pm3$~dB transition windows.}
   \label{fig:boundary}
 \end{subfigure}
 \\[2pt]
 \begin{subfigure}{\columnwidth}
    \centering
   \includegraphics[width=0.8\columnwidth]{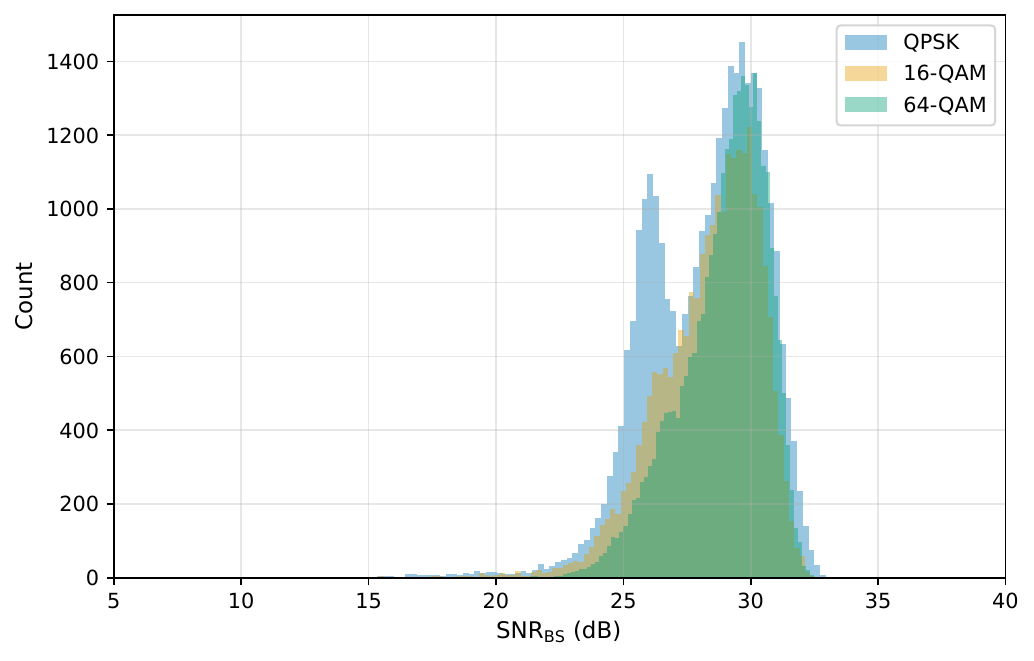}
   \caption{$\text{SNR}_{\text{BS}}$ distributions across modulation classes.}
   \label{fig:bs-snr}
 \end{subfigure}
 \caption{SNR distributions for MS-side and BS-side.}
 \label{fig:snr_distributions}
\end{figure}

The measured $\mathrm{SNR}_{\mathrm{MS}}$ spans 11.9--29.6~dB, covering the transition from unreliable to reliable operation for the considered modulation schemes. In contrast, the class-conditional $\mathrm{SNR}_{\mathrm{BS}}$ means are tightly clustered at 28.2, 28.4, and 28.9~dB for QPSK, 16-QAM, and 64-QAM, respectively, with within-class standard deviations of 1.7--2.3~dB. As shown in Fig.~\ref{fig:bs-snr}, the
$\mathrm{SNR}_{\mathrm{BS}}$ distributions strongly overlap across
modulation classes, consistent with BS-side array gain and the use of
uplink-pilot SNR while varying downlink transmit power. This limits
the discriminative value of $\mathrm{SNR}_{\mathrm{BS}}$ and
motivates the CFR-derived representation.

\subsection{Offline Classification Performance}
\label{sec:offline_classification}

\begin{table}[tb]
\centering
\caption{Classification accuracy (\%) on the full dataset and 
transition region subset. CV: 5-fold cross-validation mean $\pm$ std.}
\label{tab:accuracy_combined}
\begin{tabular}{llcccc}
\toprule
& & \multicolumn{2}{c}{Full dataset} & 
  \multicolumn{2}{c}{Transition regions} \\
\cmidrule(lr){3-4} \cmidrule(lr){5-6}
Classifier & Feature set & Test & CV & Test & CV \\
\midrule
DT  & SNR-only  & 42.0 & $41.5\pm0.1$ & 39.6 & $39.7\pm0.3$ \\
DT  & CFR-only  & 73.6 & $73.2\pm0.3$ & 78.2 & $77.8\pm0.4$ \\
DT  & CFR + SNR & 73.5 & $73.4\pm0.3$ & 78.4 & $77.7\pm0.5$ \\
\midrule
RF  & SNR-only  & 35.3 & $35.2\pm0.3$ & 35.4 & $34.8\pm0.5$ \\
RF  & CFR-only  & 81.4 & $81.2\pm0.4$ & 81.9 & $81.6\pm0.2$ \\
RF  & CFR + SNR & 81.6 & $81.6\pm0.3$ & 82.8 & $82.1\pm0.2$ \\
\midrule
KNN & SNR-only  & 37.3 & $37.2\pm0.2$ & 37.1 & $37.5\pm0.7$ \\
KNN & CFR-only  & 80.0 & $79.8\pm0.4$ & 81.1 & $80.9\pm0.2$ \\
KNN & CFR + SNR & 78.0 & $78.1\pm0.3$ & 80.7 & $79.9\pm0.3$ \\
\bottomrule
\end{tabular}
\end{table}
Table~\ref{tab:accuracy_combined} compares the performance of the three classifiers across the considered BS-side feature sets. SNR-only models perform poorly, consistent with the strong overlap in $\mathrm{SNR}_{\mathrm{BS}}$ observed in Section~\ref{sec:data}, whereas CFR-derived features substantially improve both test and cross-validation accuracy. Adding $\mathrm{SNR}_{\mathrm{BS}}$ to the CFR features provides little additional benefit, indicating that the compact CFR representation captures most of the useful discriminative information under the considered conditions. The close agreement between the test and cross-validation results further indicates stable performance across the considered data splits. RF achieves the highest offline accuracy, while DT is selected for real-time deployment because its inference can be implemented directly as compact conditional C code. 
A depth of $d=7$, corresponding to 123 leaves, is selected by 5-fold cross-validation as a suitable performance--complexity operating point.

\subsection{Feature Importance}
\label{sec:feature_importance}

\begin{figure}[tb]
 \centering 
 \includegraphics[width=\columnwidth]{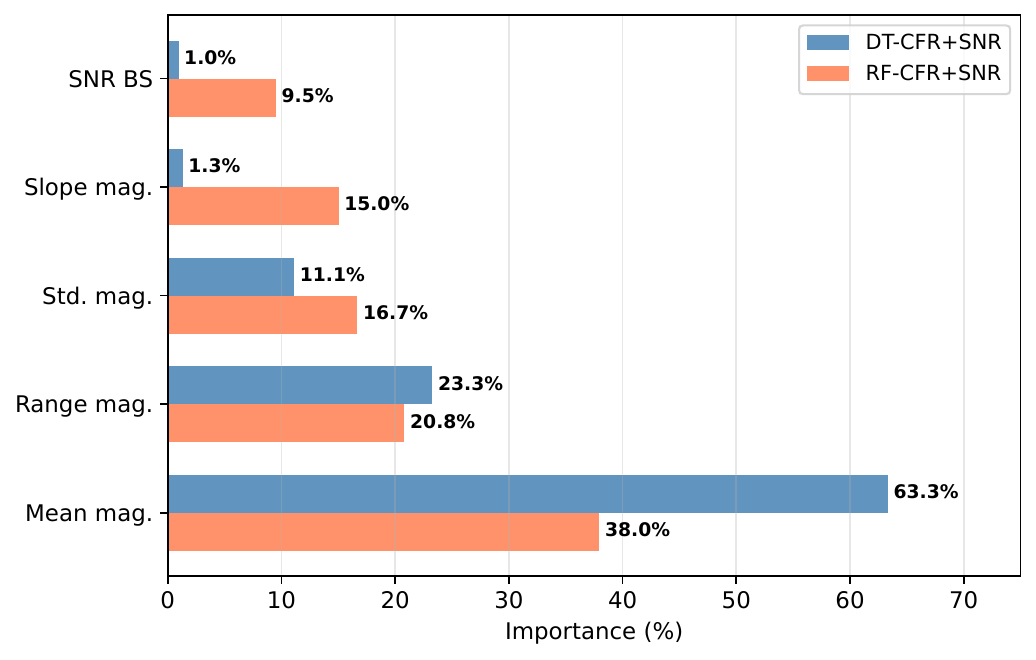}
 \caption{Feature importance for DT and RF (CFR+SNR feature subset). KNN is excluded as it does not provide a native feature importance measure. $\text{SNR}_{\text{BS}}$ contributes only 1.0\% and 9.5\% for DT and RF respectively.}
 \label{fig:feature_importance} 
\end{figure}

The classification results in Table~\ref{tab:accuracy_combined} show that the performance improvement is primarily associated with the introduction of CFR-derived features. To examine which individual measurements drive this improvement, Fig.~\ref{fig:feature_importance} reports the feature importance of the DT and RF models trained using the CFR+SNR feature set.

For both classifiers, $\mathrm{SNR}_{\mathrm{BS}}$ receives substantially lower importance than the CFR-derived features, contributing only 1.0\% for DT and 9.5\% for RF. This is consistent with the limited performance gain observed when $\mathrm{SNR}_{\mathrm{BS}}$ is added to the CFR feature set and indicates that the scalar BS-side SNR provides little additional discriminative information under the considered measurement conditions.
The mean CFR magnitude $\overline{|\mathbf{H}[k]|}$ is the most influential feature for both tree-based models, accounting for 63.3\% of the DT importance and 38.0\% of the RF importance. The remaining importance is distributed among the CFR range, standard deviation, and slope features, indicating that the classifiers exploit both the average magnitude level and, to a lesser extent, the variation of the channel response across frequency. These results suggest that the observed classification gain is driven predominantly by the compact CFR representation rather than by the available scalar $\mathrm{SNR}_{\mathrm{BS}}$ measurement.

\subsection{Transition Region Analysis}
\label{sec:boundary}

Classifier performance in the previous sections was evaluated over the full range of measured channel conditions. For link adaptation, however, particular attention is required near the reliability thresholds, where small changes in channel quality can affect the suitability of a modulation scheme. To evaluate whether the CFR-derived representation remains informative in these more critical operating regions, the analysis is restricted to samples lying within $\pm3$~dB of the 10\% BLER thresholds of QPSK and 16-QAM, located at 13.0~dB and 20.4~dB, respectively. This results in the two SNR intervals $[10.0,16.0]$~dB and $[17.4,23.4]$~dB.

The resulting subset contains 56{,}211 samples, corresponding to approximately 64\% of the full dataset, with relatively balanced transmitted-modulation proportions of 37.4\% QPSK, 31.9\% 16-QAM, and 30.8\% 64-QAM. As illustrated in Fig.~\ref{fig:boundary}, the class distributions overlap substantially within these regions, providing a more challenging setting for modulation classification.

The corresponding classification results are reported in Table~\ref{tab:accuracy_combined}. The SNR-only models remain close to the performance expected from a three-class classifier with limited discriminative information, whereas the CFR-based models maintain or improve their performance relative to the full-dataset evaluation. Adding $\mathrm{SNR}_{\mathrm{BS}}$ to the CFR representation again provides only a limited additional benefit. Overall, the transition-region results indicate that the advantage of the CFR-derived representation is retained in channel conditions close to the measured BLER reliability thresholds, where modulation decisions are particularly sensitive to changes in link quality.


\section{Conclusion}
\label{conclusion}

This paper experimentally evaluated compact BS-side CFR magnitude descriptors for ML-based modulation classification in a real-time MIMO-OFDM SDR system. Using 87{,}817 OTA observations, CFR-based features consistently outperformed the SNR-only baseline, while adding $\mathrm{SNR}_{\mathrm{BS}}$ provided little additional benefit. The same trend was retained near the measured 10\% BLER reliability thresholds. Feature-importance analysis further showed that the mean CFR magnitude is the dominant input.

Although RF achieved the highest offline performance, the depth-7 DT with 123 leaves offered a more practical performance--complexity trade-off and was successfully integrated into the LabVIEW C~Node for real-time inference at 10~Hz. The evaluation is limited to stationary indoor LoS conditions and magnitude-only CFR features. Future work will consider mobility, NLoS propagation, richer CFR/CIR features, and extension to joint modulation-and-coding adaptation.


\bibliographystyle{IEEEtran}
\bibliography{export}

\end{document}